\documentstyle[11pt,newpasp,twoside,psfig,rotate,epsf]{article}

\def\edcomment#1{\iffalse\marginpar{\raggedright\sl#1\/}\else\relax\fi}
\reversemarginpar

\def\solar{\ifmmode_{\mathord\odot}\else$_{\mathord\odot}$\fi~}

\begin{document}
\title{Millimeter VLBI and Variability in AGN Jets}
\author{Thomas P. Krichbaum, D.A. Graham, A. Witzel, J.A. Zensus}
\affil{Max-Planck-Institut f\"ur Radioastronomie, Auf dem H\"ugel 69, D-53121 Bonn, Germany}
\author{A. Greve, H. Ungerechts, M. Grewing}
\affil{Institut de Radioastronomie Millim\'etrique, 300 Rue de la Piscine, 
F-38460 St. Martin d'H\`eres, Grenoble, France}

\begin{abstract}
Millimeter-VLBI images probe as deep as never before the nuclei of AGN. VLBI at 147\,GHz
yields transatlantic fringes for the first time. Now we can begin to study the relation between
jet kinematics and spectral activity with a few ten micro-arcsecond resolution.
\end{abstract}

\section{Present Status of 3mm-VLBI}
The main advantage of VLBI observations at short millimeter wavelengths (mm-VLBI)
is the very high angular resolution and the ability to study emission regions,
which are self-absorbed at longer wavelengths. The small observing beam (up to $\sim 50$\,$\mu$as
at 86\,GHz) not only provides a spatial resolution of a few hundred to thousand
Schwarzschild radii for a typical $z>0.1$ quasar, but also facilitates to locate and 
trace the motion of moving jet components with unprecedented accuracy and at early
times after initial flux density outbursts. 

To date, global VLBI observations at 86\,GHz are
performed regularly, however, with regard to
the relatively fast motion seen in many quasars ($0.1-1$\,mas/yr),
not yet with dense enough time sampling (typically 2 observations per year).
The worldwide combination of stations in the US -- like the VLBA (presently 7 antennas equipped
with receivers) and Haystack (37\,m) -- and sensitive antennas in Europe (Effelsberg 100\,m, Pico Veleta 30\,m
Onsala 20\,m and Metsahovi 14\,m) forms the most powerful global mm-VLBI array. With a typical single
baseline detection sensitivity ($7 \sigma$) of typically $\sim 0.2$\,Jy between Effelsberg and
Pico Veleta and $\sim 0.5$\,Jy on VLBA baselines, sources with $\rm S_{\rm 86\,GHz} \geq 1-2$\,Jy
can be reliably imaged with a dynamic range of up to a few hundred. In Figure 1 we
show examples of 3\,mm-VLBI images for the quasar 3C\,345 and for BL\,Lac.
\begin{figure}[t]
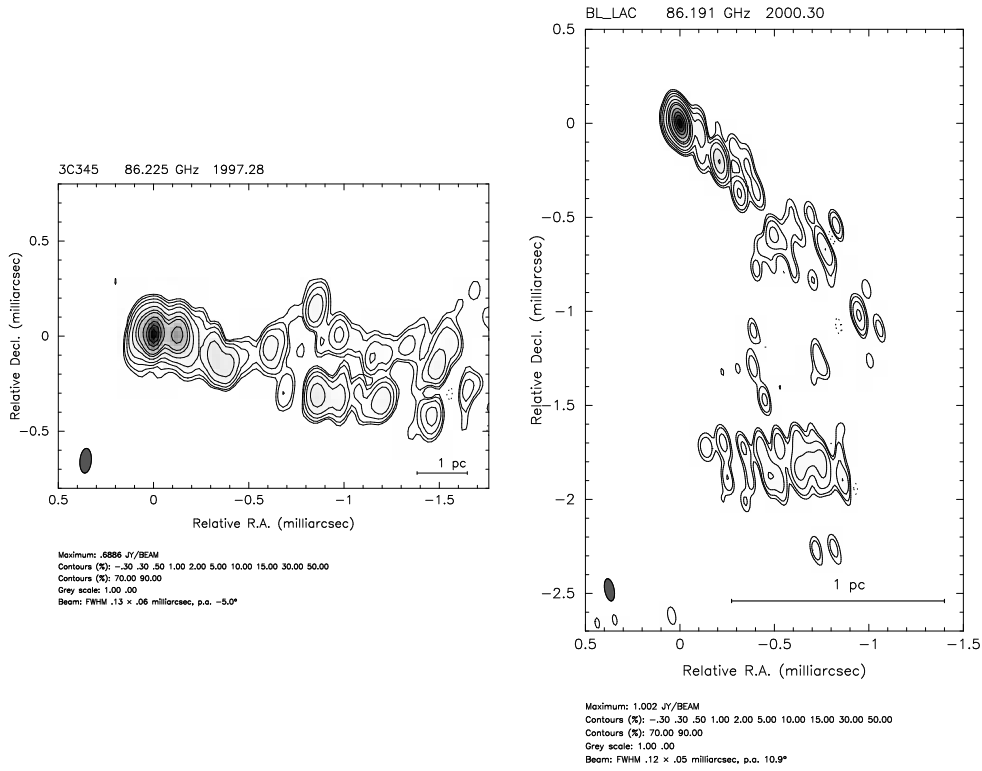

\begin{minipage}{6.5cm}{
\psfig{figure=3c345.3mm97.ps,width=6.4cm}
}
\end{minipage}
~~
\begin{minipage}{6.5cm}{
\psfig{figure=bllac.3mm00.ps,width=6.0cm}
}
\end{minipage}
\caption{Left: 3C345 at 86 GHz, observed in April 1997. The partially resolved emission 
beyond 0.5\,mas is also seen at longer wavelengths.
Right: BL\,Lac at 86 GHz, observed in April 2000. On sub-mas scales, the jet shows evidence
for bends which are followed by back-bends, indicating a helical and/or precessing jet.
} 
\end{figure}

\section{VLBI at Higher Frequencies}
Although sensitivity limitations and remaining calibration uncertainties still 
restrict the imaging capabilities of VLBI in the 3\,mm band a bit, such VLBI observations nowadays
can be regarded as more or less standard. 

Efforts to push VLBI to even shorter wavelengths started in the late 1980's and led to various
successful detections of bright AGN at 1.3\,mm, however, only on relatively short (continental) VLBI 
baselines (e.g. Greve et al. 1995, Krichbaum et al. 1997). 
Between 1995 -- 2000 several attempts with various telescopes were made to achieve fringe detections also
on the longer transatlantic baselines. These experiments were performed in the 2\,mm and 1.3\,mm bands,
but failed due to technical difficulties. The recent promising detection of 3C\,273 and 3C\,279 at 147\,GHz
on the 3100 km (1.5G$\lambda$) baseline between Pico Veleta (Spain) and Mets\"ahovi (Finland) in spring 2001 
(Greve et al. 2002), and the availability of VLBI equipment and a new 2\,mm receiver at the Heinrich
Hertz telescope (HHT) on Mt. Graham (Arizona), stimulated a new transatlantic VLBI experiment at 147\,GHz,
which was performed in April 2002. We now are able to report fringe detections also on 
the very long inter-continental 
VLBI baselines (Krichbaum et al. 2002). The detection of the 3 sources
NRAO\,150, 3C\,279, and 1633+382 on the VLBI baselines between Arizona and Spain at a fringe spacing of
4.2\,G$\lambda$ (corresponding to $\sim 49$\,$\mu$as resolution) with signal-to-noise ratios of
up to 75 (for 3C\,279) demonstrates for the first time that global VLBI at 2\,mm-wavelength is possible and that
the previous technical and sensitivity limitations can be overcome. 

The present 2\,mm and 1\,mm VLBI observations  were performed in snapshot mode (only few VLBI scans
available per source) and it is therefore not yet possible to make maps of the detected sources. However,
some estimates for the source sizes can be made. 
In Figure 2 we show the normalized visibility amplitude (=correlated flux / total flux) of 3C\,279 plotted 
versus the projected baseline length (data for April 2002: filled circles). 
The comparison of the flux density between long and short VLBI baselines
(assuming a simple Gaussian point source structure) gives a size of the VLBI
core of 3C\,279 of $\sim 35 \pm 5$\,$\mu$as at 147\,GHz. For comparison, we also show visibility points 
from the previous 2\,mm VLBI observation of 2001 (filled square) and the 1.3\,mm VLBI observation
of 1995 (open diamond). These data points indicate that the source structure is most likely
variable in time and probably not pointlike, both of which is not unexpected knowing
the source properties at longer wavelengths. 

It is possible that the factor of two 
difference between the visibility amplitudes near and below 1\,G$\lambda$ reflects
a temporal variation of the core size between $\leq 2001$ and 2002. It is 
also possible that the lower amplitude in this uv-range, is 
due to structural beating introduced by the known sub-mas jet (1\,G$\lambda$ corresponds
to 0.2\,mas). A second estimate of the source size therefore can  be made, 
if we combine the visibility points from all 3 experiments (dotted line). This 
yields a more compact source size of $\sim 25$\,$\mu$as. 

For an incoherently radiating compact synchrotron source with a brightness
temperature close to the inverse Compton limit of $10^{12}$\,K, the theoretically expected 
size is 10--20\,$\mu$as, which is about a factor of 1.5--2 smaller than the two sizes derived above. It is
therefore possible that one sees at 2\,mm wavelength the VLBI core component of 3C\,279 (the jet base) 
already spatially resolved. In this case, its intrinsic size would be of order of 0.2--0.3\,pc.
Of course, this estimate depends critically on the accuracy of the calibration of the
VLBI antennas. Future experiments with an improved gain calibration will be necessary
to check, if the rather low visibility amplitudes at G$\lambda$ uv-spacings (only $10-30$\,\% 
of the total flux density is seen) are real.

\begin{figure}
\hspace*{3cm} \psfig{figure=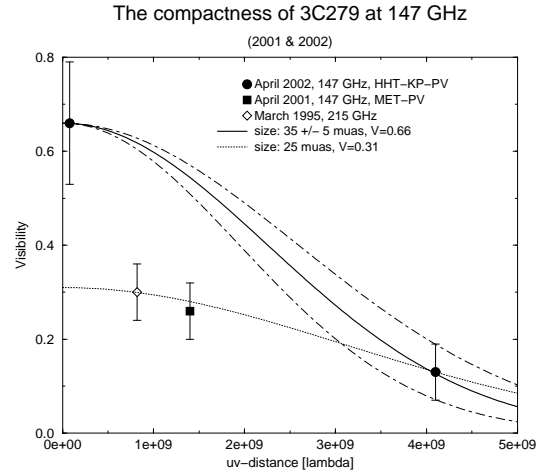,width=7cm,angle=-90}
\caption{The visibility amplitude of 3C\,279 at 147\,GHz in April 2002 (filled circles) and
in April 2001 (filled square). For comparison an earlier measurement at 215\,GHz 
is also shown (1995, Pico Veleta - Plateau de Bure). In all experiments the visibility
at $\geq 1$\,G$\lambda$ uv-spacings was lower than $V \leq 0.4$. The data indicate a size of the
VLBI core of $20-40$\,$\mu$as. For details, see text.
}
\end{figure}

\begin{figure}
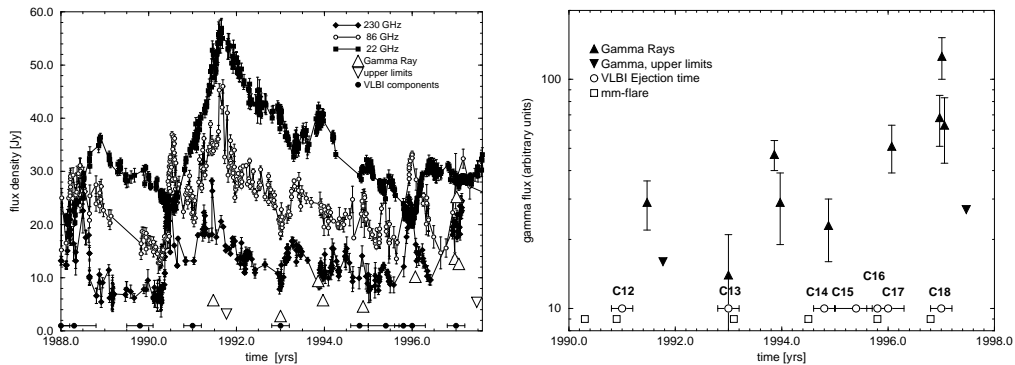

\begin{minipage}[t]{6.5cm}{
\psfig{figure=geneva4.bw.xvgr.epsi,width=6.3cm,angle=-90}
}
\end{minipage}
~~
\begin{minipage}[t]{6.5cm}{
\psfig{figure=eject.newbw.xvgr.epsi,width=6.5cm,angle=-90}
}
\end{minipage}
\caption{
Left: Flux density variations of 3C\,273 at 230 GHz (filled diamonds), 86 GHz (open circles)
and 22 GHz (filled squares). 
Upward oriented open triangles show the Gamma-ray fluxes from EGRET,
downward oriented triangles are upper limits. The extrapolated
ejection times of the VLBI components and their uncertainties
are indicated by filled circles with horizontal bars along the time axis.
Right: Times of VLBI component ejection (open circles and labels),  
times of high Gamma-ray fluxes (triangles, downward oriented symbols show upper limits) and
onset-times for millimeter flares (open squares). For details see text.
}
\end{figure}

\section{Variability and new jet components in 3C\,273}
For many AGN, a correlation between flux density variability and ejection of VLBI components 
is suggested. From our multi-frequency VLBI monitoring of 3C\,273 (1987 -- 1997, Krichbaum et al.
2002), we identified 13 jet components (C6 -- C18). These components move with apparent speeds
of 4--8\,c and seem to accelerate as they separate from the VLBI core. Our high frequency
VLBI data (15-100\,GHz) allow to trace the component motion back to the ejection from the VLBI core
and by this to determine the times of zero separation from the core.
The typical measurement uncertainty for the ejection times $t_0$ ranges between 0.2 -- 0.5\,yr. 
In Figure 3 (left), we plot $t_0$ and the millimeter-variability (22 -- 230\,GHz, data:
H. Ungerechts, H. Ter\"asranta). 
We also add the Gamma-ray detections of 3C\,273 from EGRET. 
In Figure 3 (right) we plot the onset times of the mm-flares derived
from these light-curves (T\"urler et al. 1999) together with the VLBI ejection time ($t_0$) and the 
Gamma-ray fluxes. Except for component C15, which may be not related to a flux density
outburst, a clear correlation between component ejection and onset times of mm-flares is seen.

For the Gamma-rays, the search for a correlation with mm-flares and VLBI component
ejections is complicated by the coarse time sampling of the Gamma-ray light-curve, which does
not exclude presence of more rapid variations. However, since at least 3 of the Gamma-ray detections
shown in Figure 3 (left) are followed by lower Gamma-ray fluxes (with maxima near 1991.5, 1993.9, 1997.0),
it is tempting to identify these with local maxima in a heavily undersampled light-curve.
With this assumption, it is possible to assign the ejection time of component C12 
to the Gamma-flare of 1991.5, C13 to a flare peaking
somewhere between 1993.0 and 1993.9 and C18 to the flare peaking near 1997.0.
For the Gamma-flare of 1997.0 it is very likely that the true maximum of the
light-curve has been detected, since this measurement is surrounded by two
measurements of lower intensity. For the components C14 and C15--C17 the situation is less clear.
We note, however, that the components C15--C16--C17 were ejected nearly at the same time
and are probably physically related to each other (shock-post-shock structure). Therefore,
one of the 3 components is likely to be related with the Gamma-flare peaking
around 1996.0. For C14, the ejection time correlates well with a mm-flare (peak: 1994.9) and a 
nearly contemporaneous Gamma-ray detection of moderate strength.

From a more detailed analysis
(Krichbaum et al. 2002) we obtain for the time lag between component ejection and onset of
a mm-flare: $t_0 - t_0^{\rm mm} = 0.1 \pm 0.2$\,yr. If we assume that the observed peaks in the Gamma-ray
light-curve are located near the times $t_0^\gamma$ of flux density maxima, we 
obtain  $t_0^\gamma -  t_0^{\rm mm} = 0.3 \pm 0.3$\,yr. Although the Gamma-ray variability
may be faster, this result is fully consistent with the more general finding of 
enhanced Gamma-ray fluxes mainly during the rising phase of millimeter flares.
Thus, we suggest the following sequence of events:
$t_0^{\rm mm} \leq  t_0 \leq t_0^{\gamma}$ -- the onset of a millimeter
flare is followed by the ejection of a new VLBI component and,
either simultaneously or slightly time-delayed, an increase of the Gamma-ray flux.
If we focus only on those VLBI components, which were ejected close to the main maxima of the 
Gamma-ray light-curve in Figure 3, we obtain time lags of $t_0^{\gamma} - t_0$
of $\leq 0.5$\,yr for C12, $\leq 0.9$\,yr for C13, $\leq 0.2$\,yr for
C16 and $\leq 0.1$\,yr for C18. In all cases the Gamma-rays seem to
peak a little later than the time of component ejection. With $\beta_{app} \simeq 4$
near the core, the Gamma-rays would then escape at a radius $r_\gamma \leq 0.1$\,mas.
This corresponds to $r_\gamma \leq 2000$ Schwarzschild radii (for a $10^9$ M\solar black hole) 
or $\leq 6 \cdot 10^{17}$\,cm, consistent with theoretical expectations, in which
Gamma-rays escape the horizon of photon-photon pair production at separations  
of a few hundred to a few thousand Schwarzschild radii.

\section{Summary and Future Outlook}
In summary, global mm-VLBI observations of Blazars (QSO's, BL\,Lac's) generally
show a very compact core radiating near the inverse Compton 
limit, but also a considerable ($30-50$\,\%) amount of flux and jet-like sub-structure 
on the sub-mas- to mas scale. VLBI observations at 2 \& 1\,mm indicate that the brightness
temperatures of the VLBI cores of mm-bright blazars is not significantly 
different from those, observed at longer wavelengths. The VLBI-jets often are helically bent 
and in many cases the jet curvature increases towards the VLBI-core. There is 
evidence that the helicity of the jets and the component motion along spatially curved paths 
is related to internal rotation of the jet and precession at the jet base.
In 3C\,273, 3C\,345, BL\,Lac, 0716+714 and many other sources, 
new superluminally moving jet components appear weeks to months after the $\it ONSET$ of flux
density outbursts, which propagate through the whole electromagnetic spectrum. 
In 3C\,273 Gamma-ray flares and the ejection of relativistic jet components are closely related.

The relatively fast motion and the complexity of the source structures 
requires for the future more sensitive mm-VLBI antennas, which observe with denser
(weekly to monthly) time sampling. Present day mm-VLBI suffers from the
lack of short uv-spacings. As a consequence, it is presently not possible to reliably image 
all of the source flux visible at mm-wavelengths.
The relatively high surface brightness of the sources on the short (100-1000\,km long) 
baselines facilitates easy VLBI detection and therefore gives room also for less sensitive and smaller antennas 
to play a significant role in future high resolution mm-VLBI imaging.

The addition of more collecting area through sensitive antennas specially designed for the
millimeter and sub-millimeter bands is most important for future mm-VLBI, particularly
at the shorter wavelengths (2 \& 1\,mm). This includes existing antennas, which are
not yet participating in mm-VLBI (eg. JCMT, SMA, NRO), but also new antennas like 
e.g. the 50\,m LMT in Mexico and the ALMA prototype antenna APEX.
Owing to their outstanding sensitivity, phased interferometers will play a particular important 
role. For the near future (2002/3), the participation
of the IRAM interferometer at Plateau de Bure (France) is planned (first at 3 \& 1\,mm, later
also at 2\,mm). This will increase the present sensitivity by a factor of $\sim 2-3$ to 
the $\sim 0.1$\,Jy level. In the more distant future CARMA - the merger of the BIMA and 
OVRO interferometers - and ALMA should lower the detection threshold to 1--10\,mJy and by this 
largely enhance the observational possibilities. In parallel, higher data recording rates and 
observing bandwidths (GBit/s recording using the MKV system) and correction of atmospheric
phase degradation (water vapor radiometry, phase referencing) will help to further
improve the sensitivity.

All this will facilitate the imaging of Quasars, nearby Radio-Galaxies and of the Galactic Center
source with the fascinating angular resolution of only $\sim 10-20$\,$\mu$arcseconds !

\end{document}